\documentclass[showpacs,amssymb,twocolumn,aps,nofootinbib]{revtex4}
\usepackage{amsmath}
\usepackage{amstext}
\usepackage{amsopn}
\usepackage{amsfonts}
\usepackage{amssymb}
\usepackage{bbm}
\usepackage{accents}
\usepackage{empheq}%for overbracket
\usepackage{graphicx}
\usepackage{epsf}
\usepackage{graphics}
\usepackage[latin1]{inputenc}

\begin{document}

\title{Supersymmetry, shape invariance and the hypergeometric equation}

\author{Ashok K. Das$^{a,b}$ and Pushpa Kalauni$^{c}$}
\affiliation{$^a$ Department of Physics and Astronomy, University of Rochester, Rochester, NY 14627-0171, USA}
\affiliation{$^b$ Saha Institute of Nuclear Physics, 1/AF Bidhannagar, Calcutta 700064, India}
\affiliation{$^{c}$ Instituto de Física, Universidade de São Paulo, 05508-090, São Paulo, SP, Brazil}

\begin{abstract}
It has been shown earlier \cite{bazeia} that the solubility of the Legendre
and the associated Legendre equations can be understood as a consequence of an underlying
supersymmetry and shape invariance. We have extended this result to the 
hypergeometric equation. Since the hypergeometric equation as well as the hypergeometric function reduce to various orthogonal polynomials, this study shows that the solubility of all such systems can also be understood as a consequence of an underlying supersymmetry and shape invariance. Our analysis leads naturally to closed form expressions (Rodrigues' formula) for the orthogonal polynomials.
\end{abstract}


\maketitle

\section{Introduction}

In an earlier work \cite{bazeia} it was shown that the solubility of the Legendre as well as the associated Legendre equations can be understood as a consequence of an underlying supersymmetry and shape invariance \cite{infeld, gendenshtein, cooper,das} of the differential equation. This was shown by expressing the differential operator in terms of raising and lowering operators, which not only lead to recursion relations between the polynomials naturally, but also bring out various other properties associated with the solutions. For example, this allows one to obtain a closed form expression for the Legendre polynomials (Rodrigues' formula) as well as a relation between the Legendre and the associated Legendre polynomials in a straightforward manner. 

The natural question that arises (and which was also raised in \cite{bazeia}) is whether these properties are particular to the system of Legendre equations or whether the solubility of other orthogonal polynomials can also be understood along these lines. Various attempts have already been made to understand this \cite{private}, but the question still remains open. In the present work we will study this question systematically by focusing on the hypergeometric equation. The hypergeometric equation, as well as the (Gauss') hypergeometric function 
\begin{equation}
{}_{2}F_{1} (a,b;c;z)\equiv F (a, b;c;z),\label{0}
\end{equation}
depend on three parameters (as opposed to the Legendre polynomials which depend only on one). For various values of these (three) parameters and with possible redefinition of the coordinate variable, this equation (as well as the function) reduces to those for various other orthogonal polynomials. Therefore, by choosing to study the hypergeometric equation, we are in fact studying several orthogonal polynomial equations (and polynomials) simultaneously. We will give a description of the hypergeometric equation in terms of raising and lowering operators just like the Legendre equations. This will directly lead to the underlying supersymmetry and shape invariance associated with the system. This will also allow us to determine a closed form expression (Rodrigues' formula) for the solutions (orthogonal polynomials) much like in the case of the Legendre polynomials. 

The concept of conventional supersymmetry and shape invariance has already been discussed in \cite{bazeia, gendenshtein,cooper,das}. However, as we will see shortly, the differential operator for the hypergeometric equation is non-Hermitian. The concept of supersymmetry and shape invariance can also be extended to non-Hermitian Hamiltonians (operators) \cite{levi} and here we recapitulate briefly only the essential ideas for such a case. Two Hamiltonians (operators) $BA^{\dagger}$ and $A^{\dagger}B$ correspond to supersymmetric partner Hamiltonians in the sense that 
\begin{equation}
BA^{\dagger} |\psi\rangle = E |\psi\rangle,\quad
A^{\dagger}B (A^{\dagger}|\psi\rangle) = E (A^{\dagger}|\psi\rangle),\label{0a}
\end{equation}
so that they share the same eigenvalues except for the ground state of $BA^{\dagger}$ (if the ground state energy can be scaled to zero). (Conventional supersymmetry corresponds to the case $B=A$.) While in the case of conventional supersymmetry all the energy levels are positive semi-definite, here they can be negative as well \cite{levi}. If these Hamiltonians depend on a parameter (say, a coupling constant) such that
\begin{align}
& H_{-}(a_{0}) = B(a_{0}) A^{\dagger}(a_{0}),\notag\\
& H_{+}(a_{0}) = A^{\dagger}(a_{0})B(a_{0}) = B(a_{1})A^{\dagger}(a_{1}) + C_{1},\label{0b}
\end{align}
where $a_{1}$ is a new parameter (related to $a_{0}$) and $C_{1}$ is a constant shift in the energy level (depending on the parameters), one says that the two systems (potentials) are shape invariant. In this case, we can introduce a family of Hamiltonians ($k=1,2,\cdots$)
\begin{align}
& H^{(0)} = H_{-} (a_{0}),\notag\\
& H^{(1)} = H_{-}(a_{1}) + C_{1},\cdots ,\notag\\
& H^{(k)} = H_{-}(a_{k}) + \sum_{m=1}^{k} C_{m},\cdots ,\label{0c}
\end{align}
which will be pairwise supersymmetric. Supersymmetry relates only a pair of Hamiltonians (operators). Shape invariance goes one step further and allows us to define a sequence (family) of Hamiltonians (operators) where every neighboring pair of Hamiltonians are supersymmetric. This has the consequences that the $n$th level of the original Hamiltonian can now be determined from
\begin{equation}
E_{n} = \sum_{k=1}^{n} C_{k}.\label{0d}
\end{equation}
 
Our presentation is organized as follows. In section {\bf II} we describe the hypergeometric equation and the hypergeometric function as well as its reduction to various other orthogonal polynomials for specific values of the parameters. The hypergeometric function, as we have already pointed out, depends on three parameters and, therefore, there are various ways of raising and lowering their values. In section {\bf III}, we construct the raising and lowering operators for only the first two of them  from the known recursion relations between the functions which allows us to express the hypergeometric equation as an eigenvalue problem. The raising and lowering operators allow us to determine a closed form solution for a wide range of parameter values. In section {\bf IV} we discuss the underlying supersymmetry and shape invariance associated with the system. In section {\bf V} we summarize the raising and lowering operators for other parameters as well and conclude with a brief summary.

\section{The hypergeometric equation and the hypergeometric function}

The hypergeometric differential equation can be written as the eigenvalue equation
\begin{equation}
\left(H-ab\right)F(a, b; c; z)=  0,\label{1}
\end{equation}
where we have identified
\begin{equation}
H\equiv H(a,b,c) = z(1-z)\frac{d^{2}}{dz^{2}}+(c-(a+b+1)z)\frac{d}{dz},\label{2}
\end{equation}
for simplicity. Here $a,b,c$ denote the three parameters on which the equation depends and  $F (a,b;c;z)$ is known as the (Gauss') hypergeometric function defined as 
\begin{equation}
F (a,b;c;z)=  \sum_{n=0}^{\infty}\frac{(a)_{n}(b)_{n}}{(c)_{n}}\frac{z^{n}}{n!} =\, _{2}F_{1}(b,a;c;z),\label{3}
\end{equation}
with the Pochhammer symbol given by 
\begin{equation}
(x)_{n}=  x(x+1)(x+2)\cdots (x+n-1).\label{4}
\end{equation}
We note here that the differential operator $H$ in \eqref{2} has several symmetries,
\begin{align}
& H(a,b,c) = H(b,a,c),\notag\\ 
& H(a,b,c) = H(a-n,b+n,c).\label{4a}
\end{align}
The first of the symmetries in \eqref{4a} together with the symmetry in the differential equation \eqref{1} leads to the symmetry in the hypergeometric functions \eqref{3} under $a\leftrightarrow b$. We note here that, unlike the case of the Legendre equation, the operator $H$ in \eqref{2} is not Hermitian for arbitrary values of the parameters $a,b,c$.

As we have already mentioned, the hypergeometric function $F(a,b;c;z)$ reduces to several other orthogonal polynomials for special values of the parameters. For example, we know that \cite{as,GR}
\begin{align}
& P_{n}(z)  = F (-n, n+1;1;\frac{1-z}{2}),\nonumber \\
& P_{n}^{m}(z)  =\frac{(z+1)^{\frac{m}{2}}}{(z-1)^{\frac{m}{2}}}\frac{1}{\Gamma(1-m)}\,F (-n, n+1;1-m;\frac{1-z}{2}),\nonumber \\
& T_{n}(z) = F (-n, n,\frac{1}{2};\frac{1-z}{2}),\nonumber \\
& P_{n}^{\alpha,\beta}(z) =\frac{(\alpha+1)_{n}}{n!}\,F (-n, n+\alpha+\beta+1;\alpha+1;\frac{1-z}{2}),\nonumber \\
& C_{n}^{\alpha}(z) =\frac{\Gamma(n+2\alpha)}{n!\Gamma(2\alpha)}F(-n,n+2\alpha;\alpha+\frac{1}{2};\frac{1-z}{2}),\label{5}
\end{align}
where $P_{n}(z), P_{n}^{m}(z), T_{n}(z), P_{n}^{\alpha,\beta}(z)$ 
and $C_{n}^{\alpha}(z)$ denote respectively the Legendre, associated Legendre, Chebyshev,
Jacobi and Gegenbauer polynomials. In addition there are other hypergeometric functions such as Kummer's (confluent hypergeometric) function $M(a;c;z)$, Tricomi's (confluent hypergeometric) function $U(a,b;z)$ etc. which are related to the Gauss' hypergeometric function as

\begin{align}
M(a;c;z) & = {}_{1}F_{1}(a;c;z)=\lim_{b\rightarrow\infty}{}_{2}F_{1}(a, b; c;\frac{z}{b}),\notag\\
U(a,b;z) & = z^{-a}{}_{2}F_{0}(a,1+a-b;-z^{-1})\notag\\
& = \lim_{c\rightarrow \infty} z^{-a} {}_{2}F_{1} (a, 1+a-b;c;-cz^{-1}),\label{5a}
\end{align}
namely, the confluent hypergeometric functions can be obtained as limits of the Gauss' hypergeometric functions. They also reduce to several other orthogonal polynomials. For example, the associated Laguerre polynomial $L_{n}^{m}(z)$ and the modified Bessel function $K_{n}(z)$ can be identified with \cite{as, GR} 
\begin{align}
& L_{n}^{m}(z)= \frac{(m+1)_{n}}{n!} M(-n;m+1;z),\notag\\
& K_{n}(z)=\pi^{1/2}e^{-z}(2z)^{n} U(n+\frac{1}{2},2n+1;2z).\label{5b}
\end{align}
Similarly, the Bessel function $J_{n}(z)$ can be defined as $J_{n}(z)= z^{n/2}{}_{0}F_{1} (n+1;-z^{2}/4)$. In other words, the Gauss' hypergeometric function encompasses a large number of orthogonal polynomials and by studying the underlying symmetry of the hypergeometric equation we will be studying also the behavior of the equations describing all of these polynomials.

\section{Raising and lowering operators}

To write the differential operator for the hypergeometric equation \eqref{1} and \eqref{2} as part of a supersymmetric system, we need to construct the raising and lowering operators that would reflect the recursion relations for the hypergeometric functions. As we have mentioned before, the hypergeometric function depends on three independent parameters and, correspondingly, there are a set of three  operators that raise each of the three parameters while another three which lower them. However, by looking at the orthogonal polynomials in \eqref{5}, we note that for all of them $a=-n$. Therefore, when $n$ increases by unity, $a\rightarrow a-1, b\rightarrow b+1$ while $c$ remains constant in all of them. Similarly, when $n$ decreases by unity $a\rightarrow a+1, b\rightarrow b-1$ while $c$ remains constant. Therefore, in this section, we look for a raising operator which would take $a\rightarrow a-1, b\rightarrow b+1$ simultaneously while a lowering operator which would implement $a\rightarrow a+1, b\rightarrow b-1$ (and study this case in detail). We will summarize the raising and lowering operators for the other cases in section {\bf V}. (For example, in the case of the associated Legendre polynomials, we can have $n$ fixed with $m$ changing which would correspond to $a,b$ remaining constant while $c$ changing which we will comment on later.)

We can construct such operators from the standard relations given in mathematical tables for raising and lowering of a single parameter, say $a$ (see, for example, \cite{as}), namely,
\begin{align}
& \frac{d}{dz}[z^{a} F(a, b;c;z)] =  az^{a-1} F(a+1, b; c; z),\nonumber \\
& \frac{d}{dz}[z^{c-a}(1-z)^{a+b-c} F(a, b; c; z)]\notag\\
&\quad = (c-a)z^{c-a-1}(1-z)^{a+b-c-1} F (a-1, b; c; z).\label{6}
\end{align}
Using these two relations (as well as the symmetry under $a\leftrightarrow b$ of the hypergeometric function) we can construct the desired raising and lowering operators which have the explicit forms
\begin{align}
& A^{\dagger}(a,b) = 2[z(1-z)\frac{d}{dz}+\kappa(a,b)((c-a)+z(a-b-1))],\nonumber \\
& B(a,b) = -2[z(1-z)\frac{d}{dz}+ \kappa(b,a)((c-b)+z(b-a-1))]\notag\\
&\qquad\quad = - A^{\dagger}(b,a),\label{7}
\end{align}
where we have defined (for convenience)
\begin{equation}
\kappa (a,b) = \frac{b}{1-a+b}.\label{7a}
\end{equation}
Here we have suppressed the dependence of these operators on the parameter $c$ to emphasize that they only change the values of the parameters $a$ and $b$. 

It can be checked that for $a=-n, b=n+1, c=1, z=(1-x)/2$ (see \eqref{5}), these reduce to the raising and lowering operators for the Legendre polynomials \cite{bazeia}. Furthermore, the operators in \eqref{7} can be written in a factorized form as in the case of the Legendre polynomials \cite{bazeia} (although the forms are lot more complicated in the present case). For example, $A^{\dagger}(a,b)$ takes the factorized form
\begin{equation}
A^{\dagger}(a,b) = 2 z^{-\alpha+1} (1-z)^{\beta+1}\frac{d}{dz}\,z^{\alpha} (1-z)^{-\beta},\label{7b}
\end{equation}
with
\begin{equation}
\alpha(a,b) = (c-a)\kappa(a,b),\quad \beta(a,b)= (c-b-1)\kappa(a,b),\label{7c}
\end{equation}
and similarly for $B(a,b)$. These reduce to the factorized operators of Legendre polynomials for specific values of the parameters. For completeness we note here that although $B(a,b)\neq A(a,b)$ they are related through a coordinate dependent similarity transformation,
\begin{equation}
B(a,b) = S(a,b) A(a,b) S^{-1}(a,b),\label{7d}
\end{equation}
where
\begin{equation}
S(a,b) = z^{-(\alpha(a,b)+\alpha(b,a))+1} (1-z)^{(\beta(a,b)+\beta(b,a))+1}.\label{7e}
\end{equation}
The action of $A^{\dagger}(a,b), B(a,b)$ on the hypergeometric function can be worked out to give
\begin{align}
& A^{\dagger}(a,b) F(a,b;c;z) = 2\kappa(a,b)(c-a) F(a-1, b+1;c;z),\nonumber \\
& B(a,b)F(a,b;c;z) = -2\kappa(b,a)(c-b) F(a+1, b-1;c,z).\label{8}
\end{align}

From the definition of the raising and lowering operators in \eqref{8}, it follows that  
\begin{align}
\lefteqn{B(a-1,b+1)A^{\dagger}(a,b)F(a,b;c;z)}\notag\\
& =  - 4\kappa(b+1,a-1)\kappa(a,b) (c-a) (c-b-1)F(a,b;c;z).\label{9}
\end{align}
Substituting the explicit forms of the raising and lowering operators from
\eqref{7} into \eqref{9} we obtain 
\begin{align}
&[B(a-1,b+1)A^{\dagger}(a,b)-E]F(a,b;c;z)\notag\\
&\qquad = - 4z(1-z)[H-ab] F(a,b;c;z) = 0,\label{10}
\end{align}
where we have identified
\begin{equation}
E = \frac{4(a-1)b(c-a)(c-b-1)}{(1-a+b)^{2}}.\label{10a}
\end{equation}
Namely, this shows that $F(a,b;c;z)$ satisfying the recursion relations in \eqref{8} do indeed satisfy the hypergeometric equation \eqref{1}.

We note from \eqref{8} that we can write
\begin{equation}
F(a,b;c;z)=  \frac{A^{\dagger}(a+1,b-1)}{2\kappa(a+1,b-1)(c-a-1)} F(a+1,b-1;c;z).\label{11}
\end{equation}
This relation can be iterated and, in particular, if we choose $a=-n$, then the series terminates after $n$ iterations (for $b>n, c>0$, or $b,c$ noninteger) since $F(0, b-n;c;z)=1$. As a result, from the recursion relation \eqref{7} or \eqref{7b} we can obtain a closed form expression (Rodrigues' formula) for $F(-n,b;c;z)$ as (with $b>n, c>0$ or $b,c$ noninteger)
\begin{equation}
F(-n,b;c;z) = \prod_{m=1}^{n}\frac{A^{\dagger}(-n+m,b-m)}{2\kappa(a+m,b-m)(c+n-m)},\label{12}
\end{equation}
where the product of operators is assumed to act on identity. (The product in \eqref{12} is ordered with the terms with smaller values of $m$ occurring to the left.) Since the orthogonal polynomials correspond to the value $a=-n$, this relation generates a closed form expression (Rodrigues' formula) for them for a wide range of the parameters $b,c$. Using other raising and lowering operators to be discussed in section {\bf V}, one can obtain closed form expressions for other regions of the parameter space as well.

\section{Supersymmetry and shape invariance}

As we have already pointed out, the differential operator for the hypergeometric equation \eqref{2} is not Hermitian for arbitrary values of the parameters $a,b,c$. This led to the factorization of the operator in the unconventional form $B (a-1,b+1)A^{\dagger}(a,b)$ (see also the discussion around \eqref{0a}). From the discussion in \eqref{0a} we expect that the supersymmetry partner of this Hamiltonian (operator) will be given by $A^{\dagger}(a,b)B(a-1,b+1)$. We see from the defining relations in \eqref{8} that this is indeed true
\begin{align}
& (B(a-1,b+1)A^{\dagger}(a,b) - E)|\psi\rangle = 0,\notag\\
& (A^{\dagger}(a,b)B(a-1,b+1) - E)(A^{\dagger} (a,b)|\psi\rangle) = 0,\label{13}
\end{align}
where $E$ is defined in \eqref{10a}. Furthermore, as in \eqref{10}, we can calculate explicitly and obtain
\begin{align}
& A^{\dagger}(a,b)B(a-1,b+1) - E \notag\\
&\qquad = -4z(1-z) (H - (a-1)(b+1)),\label{14}
\end{align}
so that using \eqref{8} (as well as \eqref{10}, \eqref{13} and \eqref{14}), we conclude that 
\begin{equation}
|\psi\rangle \sim F(a,b;c;z),\quad A^{\dagger}(a,b)|\psi\rangle \sim F(a-1,b+1;c;z).\label{14a}
\end{equation}
This shows that $B(a-1,b+1)A^{\dagger} (a,b)$ and $A^{\dagger}(a,b)B(a-1,b+1)$ indeed define a supersymmetric system which, through their explicit forms, is reflected in the family of hypergeometric equations. In fact, it is straightforward to check that the pair of Hamiltonians $B(a-k-1,b+k+1)A^{\dagger}(a-k,b+k)$ and $A^{\dagger}(a-k,b+k)B(a-k-1,b+k+1)$ define supersymmetric pairs for $k=0,1,2,\cdots$ in the sense that
\begin{align}
& (B(a-k-1,b+k+1) A^{\dagger}(a-k,b+k) - E_{k}) |\psi_{k}\rangle = 0,\notag\\
& (A^{\dagger}(a-k,b+k) B(a-k-1,b+k+1) - E_{k})\notag\\
& \qquad\qquad \times (A^{\dagger}(a-k,b+k) |\psi_{k}\rangle) = 0,\label{15}
\end{align}
where ($k=0,1,2,\cdots$ and $E_{0}=E$)
\begin{equation}
E_{k} = \frac{4(a-k-1)(b+k)(c-a+k)(c-b-k-1)}{(2k+1-a+b)^{2}}.\label{16}
\end{equation}

Let us next note that we can write
\begin{align}
& A^{\dagger} (a-k,b+k) B (a-k-1, b+k+1)- E_{k}\notag\\
& = B (a-k-2,b+k+2)A^{\dagger}(a-k-1,b+k+1) -E_{k+1},\label{17}
\end{align}
where $k=0,1,2,\cdots$. Namely, the supersymmetric pair of Hamiltonians (operators) in \eqref{15} are in fact, shape invariant with only a shift. This is easily understood from the fact that the hypergeometric differential equation has no potential term and, therefore, is trivially shape invariant. In fact, as a consequence, all orthogonal polynomials obtained from the hypergeometric function for special values of the parameters only have equations that are trivially shape invariant with just a shift. This is not true when the orthogonal polynomial is obtained through a coordinate dependent redefinition as in the case of the associated Legendre polynomial in \eqref{5}. However, the shape invariance, in this case, has been described in detail in \cite{bazeia} and can be applied to the other orthogonal polynomials (such as the Bessel function) related to the hypergeometric function through a coordinate dependent redefinition. 

Let us next note that if we define a family of parameters $(a_{k}=a-k, b_{k}=b+k)$, then from the (second) symmetry in \eqref{4a} we have
\begin{equation}
H(a_{k},b_{k}) = H(a-k,b+k) = H(a,b).\label{18}
\end{equation} 
This also allows us to define a sequence of hypergeometric operators ($k=0,1,2,\cdots$)
\begin{equation}
H^{(k)} = H(a_{k},b_{k}) - a_{k}b_{k} = H - (a-k)(b+k),\label{19a}
\end{equation}
where we have identified
\begin{equation}
H^{(0)} = H - ab.\label{19}
\end{equation}
We can now write the shape invariant family of operators as
\begin{align}
& \left(B(a-k-1,b+k+1)A^{\dagger}(a-k,b+k) - E_{k}\right)\notag\\
&\qquad = - 4z(1-z) H^{(k)},\quad k=0,1,2,\cdots.\label{20}
\end{align}
Furthermore, from \eqref{19a} we note that we can write
\begin{equation}
H^{(k+1)} = H^{(k)} - R_{k+1},\quad  k=0,1,2,\cdots,\label{20a}
\end{equation}
where the constant shift is identified with
\begin{equation}
R_{k+1} = a-b-(2(k+1)-1).\label{20b}
\end{equation}
This shows that the shift between the adjacent levels (as has been pointed out in \cite{bazeia}) is level dependent unlike the harmonic oscillator case and this follows from the fact that
\begin{equation}
[B(a-k-1,b+k+1), A^{\dagger}(a-k,b+k)] = -4z(1-z) R_{k+1},\label{21}
\end{equation}
namely, the commutator between the raising and lowering operators depends on the level and is no longer a constant unlike in the harmonic oscillator. (Incidentally, all of this reduces to the case of Legendre equation discussed in \cite{bazeia} for specific values of the parameters.) Therefore, the $n$th level of $H^{(0)}$ is determined to be
\begin{equation}
\sum_{k=0}^{n-1} R_{k+1} = n (a-b-n),\label{22}
\end{equation}
and recalling the definition of $H^{(0)}$ from \eqref{19} we obtain the $n$th level of the hypergeometric operator $H$ to be
\begin{equation}
ab + n(a-b-n) = (a-n)(b+n),\label{23}
\end{equation}
as it should be.

\section{The other cases}

The hypergeometric function depends on three independent parameters and in the last two sections we have analyzed in detail the behavior of the equation as well as the function when only the first two parameters change. This is because this corresponds to changing the fundamental parameter $n$ of any of the orthogonal polynomials. In this section, we will briefly summarize the other possible cases. In particular, we will indicate how the raising and lowering operators are constructed which is at the heart of this analysis.

From known properties of the hypergeometric functions as in \eqref{6}, one can easily construct the operators which raise or lower only a single parameter by unity (up to a normalization). For example, for $b\rightarrow b+1$ and $b\rightarrow b-1$, the operators are given respectively by
\begin{equation}
A^{\dagger}(b) = z \frac{d}{dz} + b,\quad B(b) = z(1-z)\frac{d}{dz} + (c-b)-az,\label{24}
\end{equation}
where we continue to denote manifestly in the operators the parameter that changes. The operators in \eqref{24} can be easily checked to satisfy (see also \eqref{10})
\begin{equation}
B(b+1)A^{\dagger}(b) - b(c-b-1) = z (H-ab).\label{25}
\end{equation}
Similarly, the operators that change $c\rightarrow c+1$ and $c\rightarrow c-1$ are given respectively by (up to a normalization)
\begin{equation}
A^{\dagger}(c) = (1-z)\frac{d}{dz} + (c-a-b),\quad B(c) = z\frac{d}{dz} + (c-1),\label{26}
\end{equation}
and these operators satisfy (see \eqref{10})
\begin{equation}
B(c+1)A^{\dagger}(c) - (c-a)(c-b) = (H-ab).\label{27}
\end{equation}

Together with the two operators given in \eqref{6}, these six operators constitute the basic raising and lowering operators for the hypergeometric functions. Any other operator can be constructed from them. For example, the operators that raise and lower both $b$ and $c$ by one unit are easily obtained to be
\begin{align}
A^{\dagger}(b,c) & = -(1-z)\frac{d}{dz} + b,\notag\\
B(b,c) & = z(1-z)\frac{d}{dz} + (c-1-az),\label{28}
\end{align}
and the two operators satisfy
\begin{equation}
B(b+1,c+1)A^{\dagger}(b,c) - b(c-a) = - (1-z) (H-ab).\label{29}
\end{equation}
The behavior of the hypergeometric equation when one (or more) of the parameters is changed can be studied exactly in the same manner as has been described in detail in the last two sections and, in fact, will lead to closed form expressions for the orthogonal polynomials in different regions of the parameter space as has been noted earlier.

To conclude, we have generalized the analysis of \cite{bazeia} to the hypergeometric equation which depends on three independent parameters and encompasses equations for several orthogonal polynomials. This shows that the solubility of various orthogonal equations can also be understood as due to an underlying supersymmetry and shape invariance in the system. The analysis also yields a closed form expression (Rodrigues' formula) for the orthogonal polynomials exactly in the same way as in the case of the Legendre polynomials.

A. D. would like to thank USP for hospitality and support while P. K. acknowledges FAPESP (Brazil) - Process No. 2013/08090-9 for financial support.

\end{document}